\documentclass[twocolumn]{revtex4}
\usepackage{graphicx}

\newcommand{\be}{\begin{equation}}
\newcommand{\ee}{\end{equation}}
\newcommand{\ba}{\begin{eqnarray}}
\newcommand{\ea}{\end{eqnarray}}
\newcommand{\go}{\agt}
\newcommand{\lo}{\alt}

\def\bE{{\bf E}}
\def\bB{{\bf B}}
\def\GeV1{{\rm GeV}^{-1}}
\def\hatq{{\hat q}}

\begin{document}
\title{Probing Axions with Radiation from Magnetic Stars}
\author{Dong Lai}
\email{dong@astro.cornell.edu}
\affiliation{Center for Radiophysics and Space Research,
Department of Astronomy, Cornell University, Ithaca, NY 14853}
\author{Jeremy Heyl}
\email{heyl@phas.ubc.ca}
\affiliation{Department of Physics and Astronomy,
University of British Columbia, Vancouver BC V6T 1Z1 Canada; Canada Research Chair}

\begin{abstract}
  Recent experiments suggest that polarized photons may couple
  significantly to pseudoscalar particles such as axions.  We study
  the possible observational signatures of axion-photon coupling for
  radiation from magnetic stars, with particular focus on neutron
  stars. We present general methods for calculating the axion-photon
  conversion probability during propagation through a varying
  magnetized vacuum as well as across an inhomogeneous
  atmosphere. Partial axion-photon conversion may take place in the
  vacuum region outside the 
neutron star. Strong axion-photon mixing occurs
  due to a resonance in the atmosphere, and depending on the axion
  coupling strength and other parameters, significant axion-photon
  conversion can take place at the resonance. Such conversions may
  produce observable effects on the radiation spectra and polarization
  signals from the star.  We also apply our results to axion-photon
  propagation in the Sun and in magnetic white dwarfs. 
We find that there is no appreciable conversion of solar axions
to photons during the propagation.
\end{abstract}
\maketitle

\section{Introduction}

The axion is a hypothesized pseudoscalar particle arising from the
breaking of a $U(1)$ Peccei-Quinn symmetry, introduced to explain 
the absence of strong CP violation\cite{peccei,weinberg,wilczek}.
The axion is also an ideal candidate for cold dark matter that makes up 
five-sixths of all matter in the universe. The allowed axion mass
$m_a$ is in the range of $10^{-6}\lo m_a\lo 10^{-3}$~eV\cite{raffelt04}.

A general property of the axion is that it can couple to two photons
(real or virtual) via the interaction \be {\cal
  L}_{a\gamma\gamma}=-{1\over 4}\,g\,a\,F_{\mu\nu}{\tilde F}^{\mu\nu}=
g\,a\,\bE\cdot\bB, \ee where $a$ is the axion field, $F_{\mu\nu}$
(${\tilde F}^{\mu\nu}$) is the (dual) electromagnetic field strength
tensor, and $g$ is the photon-axion coupling constant. Accordingly, in
the presence of a magnetic field ${\bf B}$, a photon may oscillate
into an axion and vice versa.  Exploiting such photon-axion
oscillation, experiments based on ``photon regeneration'' (``invisible
light shining through walls'') set the constraint $g<6.7\times
10^{-7}~\GeV1$ ($95\%$ CL) for axions with
$m_a<10^{-3}$~eV\cite{cameron93}. The first results from the CERN
Axion Solar Telescope (CAST) experiment imply an upper limit $g\lo
10^{-10}~\GeV1$ for $m_a\lo 0.02$~eV \cite{zioutas05}. Other
experimental and astrophysical constraints on the axion parameters
(mass $m_a$ and $g$) are reviewed in
Refs.~\cite{raffelt04,raffelt05,vanBibber06}.

Recently, the PVLAS collaboration has reported measurements of the 
rotation of the polarization of photons in passing through a vacuum cavity
in a magnetic field~\cite{zavattini06}. If interpreted as due to the 
conversion of photons to axions, this may imply $g\simeq (1.6-5)
\times 10^{-6}~\GeV1$ for $m_a\sim 10^{-3}$~eV. Unfortunately, this result 
contradicts the constraint from the CAST experiment and 
the previous energy-loss limit from globular cluster stars.
Thus, either there remain systematic effects in the PVLAS experiment
or nonminimal models of psuedoscalar-photon coupling are required
\cite{masso05}.

In this paper, we study the effects of axion-photon coupling on
radiation from magnetic stars (see \cite{raffelt88} for previous
works).  We will consider generic psuedoscalar parameters with $m_a\lo
10^{-3}$~eV and $g\lo 10^{-6}~{\rm GeV}^{-1}$, 
not necessarily the QCD axion 
(for which there exists a unique
relation between $m_a$ and $g$).
Our goal is to understand and quantify what might be the
observational signatures of axion-photon coupling for various axion
parameters.  Our method and basic results (such as the axion-photon
conversion probability) can be applied for all axion parameters and in
different astrophysical environments. Most of the paper will deal with
magnetic neutron stars, for which most significant effects are
possible, but we also discuss applications of our results in the
contexts of magnetic white dwarfs and the Sun.

Our paper is organized as follows. In Sect.~II we study axion-photon
propagation in the vacuum region outside a magnetic neutron star,
taking into account of the variation of the magnetic field.  Section
III examines the propagation in a magnetized plasma characteristic of
a neutron star atmosphere. We study in detail the properties of the
axion-photon resonance as the photon-axion propagates in an
inhomogeneous medium. Depending on the axion coupling strength and
other parameters, significant axion-photon conversion may take place
and such conversion may leave an imprint on the radiation spectrum and
polarization from the star.  In Sect.~IV we consider axion-photon
propagation in the Sun and in magnetic white dwarfs, and Sect.~V
contains a brief discussion of our results.

\section{Axion-Photon Propagation in an Inhomogeneous Magnetized
  Vacuum}
\label{sec:vacuum}

Here we consider the effect of axion-photon coupling as the photon
propagates radially outwards through the vacuum region outside a magnetized 
neutron star. The stellar magnetic field is assumed to be dipolar.  
Both of these assumptions are valid several radii away from the star.

\subsection{Basic Equations}

In an external static magnetic field $\bB$, the evolution of the
photon electric field $\bE$ and the axion field $a$ of a given energy
$\omega$ (so that $\bE,a\propto e^{i\omega t}$) takes the
form~\cite{raffelt88} (in units with $\hbar=c=1$):
\be i{d\over dz}\left(\begin{array}{c}a\\
    E_\parallel\end{array}\right) =\left(\begin{array}{cc}
    \omega+\Delta_a & \Delta_M \\
    \Delta_M & \omega+\Delta_\parallel\end{array} \right)
\left(\begin{array}{c}a \\ E_\parallel \end{array}\right),
\label{eq:evol}\ee
where $E_\parallel$ is the photon electric field in the plane spanned by
$\bB$ and the $z$-axis, and we have assumed that the wave propagates in the 
$z$-direction. The matrix element $\Delta_\parallel$ arises from vacuum
polarization~\cite{heisenbergeuler36,schwinger51,adler71,heylhernquist97} 
and is given by (for $\omega\ll m_e$, the electron rest mass)
\be
\Delta_\parallel={1\over 2}\,q\omega\sin^2\theta,
\label{eq:delatapara}\ee
where $\theta$ is the angle between the direction of propagation and the 
magnetic field, and $q$ is a dimensionless function of $b=B/B_Q$,
with $B_Q=m_e^2c^3/(e\hbar)=4.414\times 10^{13}$~G the critical QED 
field strength. For $b\ll 1$, $q=7\alpha b^2/(45\pi)$. A general fitting
formula (accurate to within $3\%$ for all $b$'s) which has the correct 
$b\ll 1$ and $b\gg 1$ limits is~\cite{potekhin04}
\be
q={7\alpha \over 45\pi}b^2 {\hat q},\quad {\rm with}~~
{\hat q}={1+1.2 b\over 1+1.33 b+0.56 b^2}.
\label{eq:q}\ee
Thus
\be
\Delta_\parallel=
{7\alpha\over 90\pi}\hatq\beta^2\omega=
0.1807\,\hatq\,\beta^2\,\omega_1~{\rm eV}=9173\,\hatq\,\beta^2\omega_1~
{\rm cm}^{-1},
\ee
where $\omega_1=\omega/(1~{\rm keV})$ and
\be
\beta=b\sin\theta={B\over B_Q}\sin\theta.
\ee
The matrix element $\Delta_a$ in Eq.~(\ref{eq:evol}) is due to the finite 
axion mass, and $\Delta_M$ arises from the photon-axion coupling:
\ba
&&\Delta_a=-{m_a^2\over 2\omega}\nonumber\\
&&\qquad =-5 \times 10^{-14}\,m_5^2\omega_1^{-1}\,{\rm eV}\nonumber\\
&&\qquad =-2.538 \times 10^{-9} m_5^2\omega_1^{-1}{\rm cm}^{-1},\\
&&\Delta_M={1\over 2}gB\sin\theta={g\,m_e^2\over 2}\,\beta\nonumber\\
&&\qquad =1.306 \times 10^{-7} \,g_9\,\beta ~{\rm eV}\nonumber\\
&&\qquad =
6.627\times 10^{-3}\,g_9\, \beta\,\, {\rm cm}^{-1},
\ea
where $m_5=m_a/(10^{-5}~{\rm eV})$ and $g_9=g/(10^{-9}~{\rm GeV}^{-1})$.

Note that Eq.~(\ref{eq:evol}) is valid in the weak-dispersion limit,
i.e., $\Delta_\parallel/\omega,~|\Delta_a|/\omega$ and $\Delta_M/\omega$
are all much less than unity.

In the absence of photon-axion coupling ($\Delta_M=0$), and with 
$E_\parallel,a\propto 
e^{-ikz}$, we find $k/\omega=1+\Delta_\parallel/\omega$ and
$1+\Delta_a/\omega$ for photon (the parallel-mode) and axion, respectively.

\subsection{Evolution in a Varying Magnetic Field}
\label{subsec:evol}

It is convenient to introduce the quantity $\Delta k$ and the mixing angle
$\theta_m$ via
\be
\Delta k=2\left[(\Delta_a-\Delta_\parallel)^2/4+\Delta_M^2\right]^{1/2},
\label{eq:delk}
\ee
and 
\be
\tan 2\theta_m={\Delta_M\over (\Delta_a-\Delta_\parallel)/2}\equiv\Lambda.
\label{eq:tan}\ee
Then the evolution equation (\ref{eq:evol}) can be written as
\ba
&& i{d\over dz}\left(\begin{array}{c} a\\ E_\parallel\end{array}\right)
=\Biggl
[\left(\omega+{\Delta_\parallel+\Delta_a\over 2}\right){\bf I}\nonumber\\
&& \qquad +{\Delta k\over 2}\left(\begin{array}{cc}
\cos 2\theta_m & \sin 2\theta_m \\
\sin 2\theta_m & -\cos 2\theta_m\end{array} \right)
\Biggr]\left(\begin{array}{c} a \\ E_\parallel \end{array}\right),
\label{eq:evol2}\ea
where ${\bf I}$ is the unit $2\times 2$ matrix.

The eigenvalues and eigenvectors of Eq.~(\ref{eq:evol2}) are
\be
k_+=\omega+{\Delta_\parallel+\Delta_a\over 2}+{\Delta k\over 2},\quad
\left(\begin{array}{c} a \\ E_\parallel\end{array}\right)_+
=\left(\begin{array}{c}\cos\theta_m\\ \sin\theta_m\end{array}\right)
\ee
for the ``$+$'' mode, and 
\be
k_-=\omega+{\Delta_\parallel+\Delta_a\over 2}-{\Delta k\over 2},\quad
\left(\begin{array}{c}a \\ E_\parallel\end{array}\right)_-
=\left(\begin{array}{c}-\sin\theta_m\\ \cos\theta_m\end{array}\right)
\ee
for the ``$-$'' mode. 

\begin{figure}
\includegraphics[height=7cm]{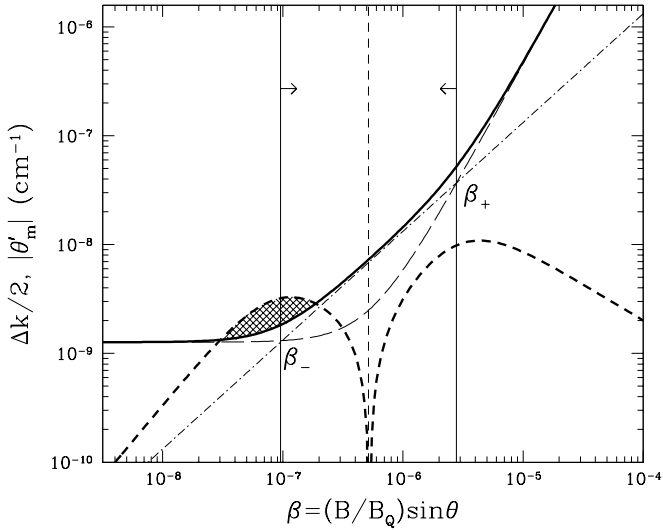}
\caption{ The solid curve shows $\Delta k/2$, and the dashed curve
  shows $|\theta_m'|$ (assuming that the field is dipolar and that at
  the stellar surface $\beta=\beta_s=1$ and $R=10^6$~cm). The light
  dot-dashed line shows $\Delta_M$ and the long-dashed line shows
  $(\Delta_\parallel-\Delta_a)/2$. The parameters are:
  $m_a=10^{-5}$~eV, $g=2\times 10^{-9}$~GeV$^{-1}$,
  $\omega=1$~keV. The ``Strong mixing'' region (between the two vertical 
  solid lines) is where
  $(\Delta_\parallel-\Delta_a)/2 <\Delta_M$ or $\beta_- <\beta
  <\beta_+$; nonadiabatic mode evolution occurs in the 
  region where
  $\Delta k/2 <|\theta_m'|$ (cross-hatched area).  The vertical dashed line denotes $\beta_*$ where the maximum mixing angle is achieved.
  \label{fig:f1}}
\end{figure}

Figure \ref{fig:f1}
shows an example of the mode eigenvalue as a function of $\beta$
(with the other parameters fixed). For sufficiently
large $\beta$ (such that $\Delta_\parallel\gg \Delta_M,|\Delta_a|$), 
we have $\Delta k\simeq \Delta_\parallel-\Delta_a$, thus $k_+=
\omega+\Delta_\parallel$, $k_-\simeq\omega+\Delta_a$ and 
$\tan 2\theta_m\simeq 0$. As $\beta$ decreases, $|\tan 2\theta_m|$ 
first increases
and then decreases. For sufficiently small $\beta$ (such that 
$|\Delta_a|\gg \Delta_M,~\Delta_\parallel$), we again have
$\Delta k\simeq \Delta_\parallel-\Delta_a$, $k_+=
\omega+\Delta_\parallel$, $k_-\simeq\omega+\Delta_a$ and $\tan 2\theta_m
\simeq 0$.
Thus, for both large and small $\beta$, 
the ``$+$'' mode always represents the axion and 
the ``$-$'' mode always represents the photon. Strong mixing may occur 
for intermediate $\beta$'s when $\tan 2\theta_m\go 1$, but there is no
``level crossing''.

Consider a photon created in the large-$\beta$ region of a neutron star. As
the photon propagates outward to the low-$\beta$ region, 
it may partially convert to the axion.
To calculate the net conversion probability, we define 
the mode amplitudes to be $A_+$ and $A_-$, and expand the
photon-axion system in terms of the normal modes:
\be
\left(\begin{array}{c}a \\ E_\parallel\end{array}\right)
=A_+\left(\begin{array}{c}\cos\theta_m\\ \sin\theta_m\end{array}\right)
+A_-\left(\begin{array}{c}-\sin\theta_m\\ \cos\theta_m\end{array}\right).
\ee
Submitting the above in Eq.~(\ref{eq:evol2}), we find that the 
mode amplitudes satisfy the equation
\be
i{d\over dz}\left(\begin{array}{c}A_+\\ A_-\end{array}\right)
=\left(\begin{array}{cc}
\Delta k/2 & i\theta_m' \\
-i\theta_m' & -\Delta k/2\end{array} \right)
\left(\begin{array}{c} A_+ \\ A_- \end{array}\right),
\label{eq:evol3}\ee
where $'$ denotes derivative with respect to $z$. In deriving
Eq.~(\ref{eq:evol3}) we have neglected the non-essential diagonal term
(proportional to ${\bf I}$) in Eq.~(\ref{eq:evol2}).

In deriving the evolution equation (Eq.~\ref{eq:evol3}) we have
assumed that the orientation of plane spanning the photon wave number
${\bf k}$
and the magnetic field ${\bf B}$ does not change along the ray path.  This is
valid for a static dipole field far away from the star (more than a
few stellar radii away) and well inside the light cylinder radius
($c/\Omega_s$, where $\Omega_s$ is the rotation rate of the star).
For a rotating star, the orientation
of the ${\bf k}$-${\bf B}$ plane changes along the ray path, but this is 
a rather small effect if the rotation rate is much less than the maximum
(breakup) value. Note that the static dipole approximation
breaks down for $r$ beyond the light cyclinder radius.

The mode evolution will be adiabatic if the condition 
\be
\gamma\equiv {\Delta k/2\over |\theta_m'|}\gg 1
\label{eq:gamma}
\ee
is satisfied. From Eq.~(\ref{eq:tan}), we find 
\be
\theta_m'={\sin 4\theta_m\over 4}{\Lambda'\over \Lambda},\quad
{\rm with}~~
{\Lambda'\over \Lambda}\simeq -{\beta'\over\beta}{\Delta_a+\Delta_\parallel
\over\Delta_a-\Delta_\parallel},
\ee
where in the $\Lambda'/\Lambda$ expression we have set $\hatq'\simeq 0$
for simplicity. For a dipolar field, we have 
\be
\beta\simeq \beta_s(R/r)^3,
\ee
where $\beta_s$ is the field strength at the stellar surface, $R=10^6R_6$~cm 
is the neutron star 
radius. Thus $\beta'/\beta \simeq -3/r=-(3/R)(\beta/\beta_s)^{1/3}$.
If the adiabatic condition (\ref{eq:gamma}) is satisfied at every point along
the photon's path, then the photon will traverse the region of varying $\bB$
without any net conversion to the axion; otherwise, the photon may experience
conversion and the emergent photon intensity will be smaller than the emitted
intensity.

It is important to map out the parameter regimes for which net nonadiabatic
photon-axion conversion may occur. To this end, we note to have any appreciable
conversion, the photon must pass through a ``strong mixing'' region, where
$|\tan 2\theta_m|$ is not much less than unity. Indeed we see from 
Eq.~(\ref{eq:evol2}) that if $|\tan 2\theta_m|\ll 1$ along the photon path, 
then $E_\parallel \propto \exp(-i\int dz\,k_+)$ and no conversion occurs
regardless of whether the condition (\ref{eq:gamma}) is satisfied or not.
We define the ``strong mixing'' region by $|\tan 2\theta_m|\ge 1$,
or $\beta_- \le \beta\le\beta_+$ (see Fig.~\ref{fig:f1}), where 
\be
\beta_\pm={45\pi g m_e^2\over 7\alpha\omega}\left[1\pm
\sqrt{1-{7\alpha m_a^2\over 45\pi (gm_e^2)^2}}\right].
\ee
In the above equation, we have used $\hatq=1$ since for the parameters
regime of interest, $\beta_\pm\ll 1$ is well satisfied.
Thus the ``strong mixing'' region exists only when
\ba
&& \qquad ~{gm_e^2\over m_a}\ge 
\left({gm_e^2\over m_a}\right)_\ast=\left (\frac{7\alpha}{45\pi}\right)^{1/2}, \nonumber\\
&&
\quad {\rm or}\quad 
{g_9\over m_5}\ge \left({g_9\over m_5}\right)_\ast=0.728
\label{eq:gm}
\ea
At the threshold point, $\beta_\pm$ collapse to 
\be
\beta_\ast={m_a\over\omega}\sqrt{45\pi\over 7\alpha}=52.6{m_a\over \omega}
=5.26\times 10^{-7}\,\frac{m_5}{\omega_1},
\ee
and 
$(\Delta k/2)_\ast=m_a^2/(\sqrt{2}\omega)$.

In general (not only at the threshold point) the maximum mixing angle
occurs where $\beta=\beta_\ast$ and has the value 
\be 
|\tan 2\theta_m|_{\rm max}
= \frac{g_9}{m_5} \left ( \frac{g_9}{m_5} \right
)_\ast^{-1} 
\ee

\begin{figure}
\includegraphics[height=7cm]{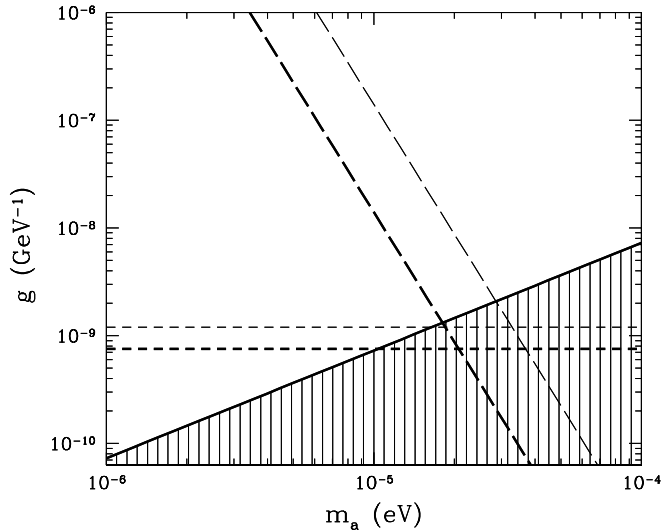}
\caption{
The parameter domain for which nonadiabatic photon-axion conversion is 
possible. Above the solid line [Eq.~(\ref{eq:gm})] (the unshaded regime),
there exists a ``strong mixing'' zone ($|\tan 2\theta_m|>1$) between photons
and axions. In the unshaded region, 
we have $g_1/m_1>(g_1/m_1)_*$, and
well inside the unshaded region the asymptotic results of
Eqs.~(\ref{eq:g1+}) and (\ref{eq:g1-}) apply.  In this (unshaded) region
and below the long-dashed line, mode evolution is nonadiabatic at 
$\beta=\beta_-$ [Eq.~(\ref{eq:g1-})];
below the short-dashed line, the evolution also becomes nonadiabatic
at $\beta=\beta_+$ [Eq.~(\ref{eq:g1+})].
The heavy lines are for $f=\omega_1^2\beta_s^{-1}R_6^{-3}
=1$, and lighter lines for $f=10$.
\label{fig:f2}}
\end{figure}

Suppose $g/m_a> (g/m_a)_\ast$, then the region where the adiabatic
condition (\ref{eq:gamma}) is most likely to be violated is around
$\beta=\beta_\pm$, where $|\theta_m|=\pi/8$ and $|\sin 4\theta_m|=1$
(see Fig.~1).  First consider the the parameter regime $g/m_a\gg
(g/m_a)_\ast$.  In this case, $\beta_+$ is determined by
$\Delta_M\simeq \Delta_\parallel/2\gg |\Delta_a|$, which gives
$\beta_+\simeq 90\pi gm_e^2/(7\alpha\omega)$. 
The nonadiabaticity condition $\Delta k/2 <|\theta_m'|$ at
$\beta=\beta_+$ then translates to 
\be 
g_9<0.755 (\omega_1^2\beta_s^{-1}R_6^{-3})^{1/5}.
\label{eq:g1+}
\ee
At $\beta=\beta_-$, which is determined by $\Delta_M\simeq |\Delta_a/2|
\gg |\Delta_\parallel|$, giving
$\beta_-\simeq m_a^3/(2\omega gm_e^2)
=1.9\times 10^{-6}m_1^2g_1^{-1}\omega_1^{-1}$, 
the nonadiabaticity condition $\Delta k/2
<|\theta_m'|$ translates to 
\be
g_9<14\, m_5^{-4}(\omega_1^2\beta_s^{-1}R_6^{-3}).
\label{eq:g1-}
\ee
Figure~\ref{fig:f2} illustrates the parameter domain for which nonadiabatic
photon-axion conversion may be possible.

\subsection{Numerical Calculations of Photon-Axion Conversion 
Probability}

To obtain the photon-axion conversion probability, it is necessary to 
perform numerical calculations. 
To this end, we rewrite Eq.~(\ref{eq:evol3}) as
\be
i{d\over d\ln\beta}\left(\begin{array}{c}A_+\\ A_-\end{array}\right)
=\left(\begin{array}{cc}
C  & iD \\
-iD  & -C\end{array} \right)
\left(\begin{array}{c} A_+ \\ A_- \end{array}\right),
\label{eq:evol4}\ee
where 
\be
C={\Delta k\over 2}{dr\over d\ln\beta},\quad
D={d\theta_m\over d\ln\beta}.
\ee
where we have assumed that $\beta$
decreases monotonically along the photon path; this is appropriate
more than a few stellar radii from the stellar surface.
At large $\beta$, we consider a pure photon state, with $A_-=1$ and $A_+=0$.
We integrate Eq.~(\ref{eq:evol4}) toward small $\beta$ and obtain the 
asymptotic $A_+(\beta\rightarrow 0)$, and the net conversion probability is
then $P_{\rm conv}=|A_+|^2$.

\begin{figure}
\includegraphics[height=9cm]{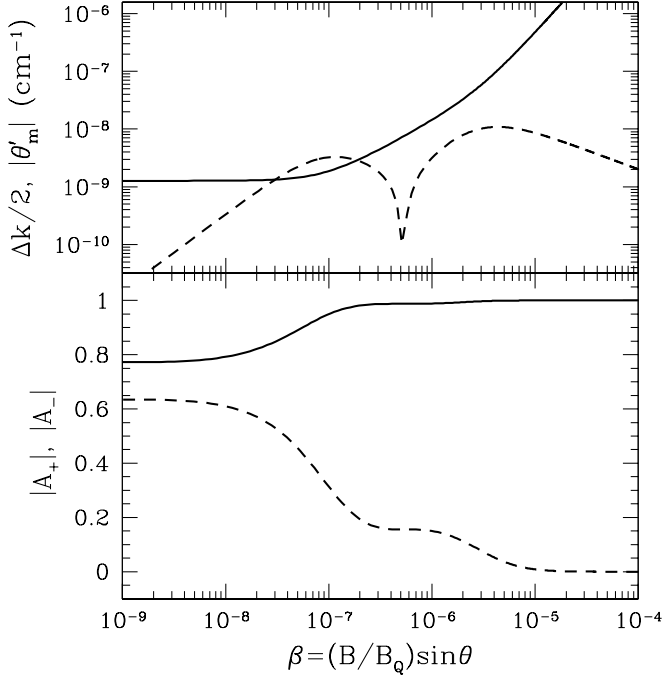}
\caption{The upper panel shows $\Delta k/2$ (solid line) and
$|\theta_m'|$ (dashed line), similar to Fig.~1.
The lower panel shows the evolution of mode amplitude, starting from
$A_-=1,~A_+=0$ at a small radius (large $\beta$). The parameters 
are $m_a=10^{-5}$~eV, $g=2\times 10^{-9}$~GeV$^{-1}$, $\omega=1$~keV,
$\beta_s=1$, and $R_6=1$.
\label{fig:evo1}}
\end{figure}

\begin{figure}
\includegraphics[height=9cm]{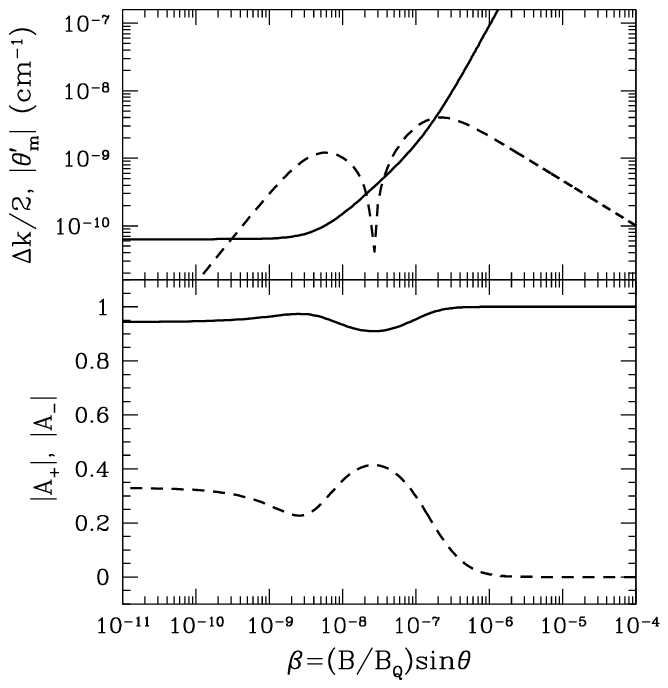}
\caption{Same as Fig.~3 except $\omega=20$~keV.
\label{fig:evo2}}
\end{figure}

\begin{figure}
\includegraphics[height=9cm]{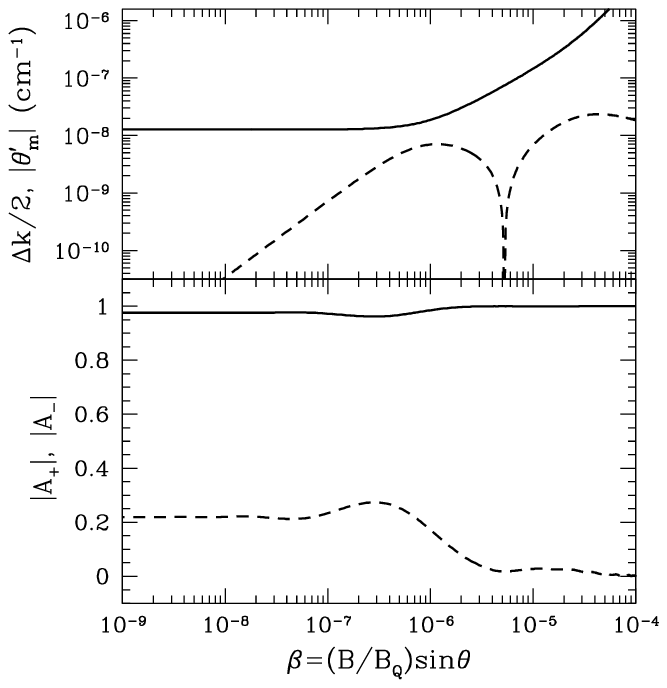}
\caption{Same as Fig.~3 except $\omega=0.1$~keV.
\label{fig:evo3}}
\end{figure}

\begin{figure}
\includegraphics[height=9cm]{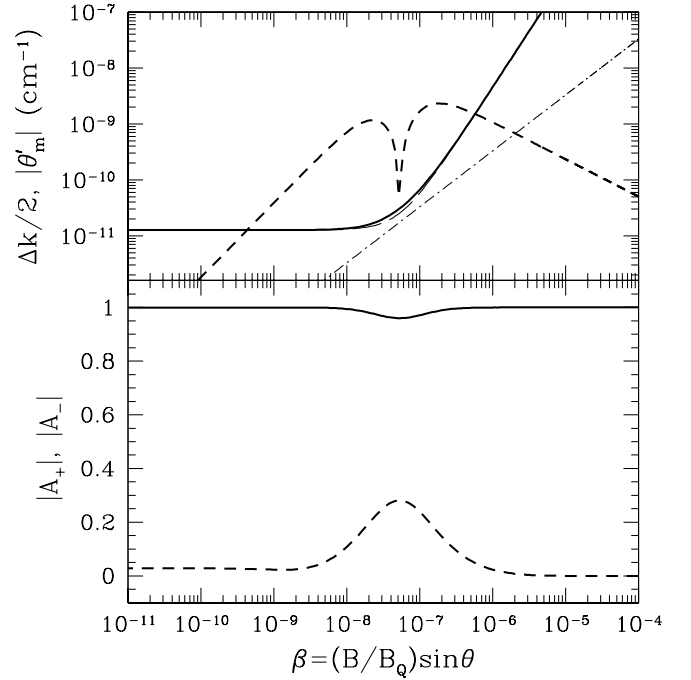}
\caption{Same as Fig.~3 except 
$m_a=10^{-6}$~eV, $g=5\times 10^{-11}$~GeV$^{-1}$, $\omega=1$~keV,
$\beta_s=1$, and $R_6=1$. In the upper panel, 
the light dot-dashed line
shows $\Delta_M$ and the long-dashed line shows 
$(\Delta_\parallel-\Delta_a)/2$. Unlike the cases shown in 
Figs.~(\ref{fig:evo1})-(\ref{fig:evo3}), no ``strong mixing'' region 
exists in this case.
\label{fig:evo4}}
\end{figure}

\begin{figure}
\includegraphics[height=7cm]{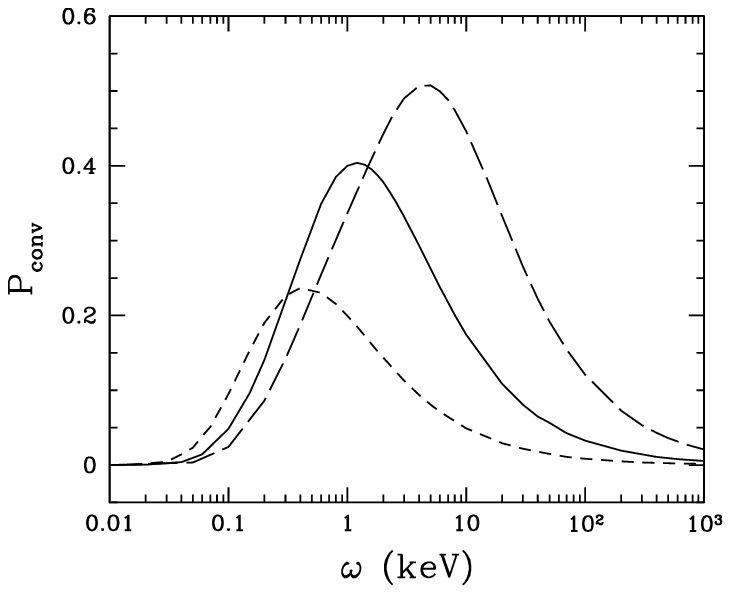}
\caption{Photon-axion conversion probability $P_{\rm conv}$ 
in vacuum as a function of photon energy. The parameters are
$m_a=10^{-5}$~eV, $\beta_s=1$, $R_6=1$; 
the solid line is for $g=2\times 10^{-9}$~GeV$^{-1}$, 
the short-dashed line for $g=10^{-9}$~GeV$^{-1}$, and
the long-dashed line for $g=4\times 10^{-9}$~GeV$^{-1}$.
\label{fig:evodat}}
\end{figure}

Figures \ref{fig:evo1}-\ref{fig:evo4} give several examples of our
numerical integrations.  Figures \ref{fig:evo1}-\ref{fig:evo3}
consider the regime $(g/m_a)>(g/m_a)_\ast$.  At low energy
(Fig.~\ref{fig:evo3}), the mode does not evolve through a
nonadiabatic zone [neither (\ref{eq:g1+}) nor (\ref{eq:g1-}) are
satisfied], thus the mode conversion probability is small.  At
intermediate energy (Fig.~\ref{fig:evo1}), the condition
(\ref{eq:g1-}) is satisfied while (\ref{eq:g1+}) is not, we obtain a
large $P_{\rm conv}$. 
At large energy (Fig.~\ref{fig:evo2}), both (\ref{eq:g1-}) and
(\ref{eq:g1+}) are satisfied, and yet the net conversion probability
is small.  Fig.~\ref{fig:evo2} gives some hints as to what is
happening.  When the wave travels through a non-adiabatic zone, the
amplitudes of the modes change.  There can be two nonadiabatic zones,
one near $\beta_+$ and one near $\beta_-$. However, the sign of
$\theta_m'$ is different in these two zones; therefore, the sign of
the mixing term $D$ in Eq.~(\ref{eq:evol4}) differs as well.  When both
zones are active 
(as for $\omega=20$~keV in Fig.~\ref{fig:evo2}),
some of the changes in the
inner zone ($\beta_+$) are undone in the outer zone ($\beta_-$).  With
this in mind, the peak conversion as a function of energy $\omega$ should
occur approximately where 
Eq.~(\ref{eq:g1+}) holds but Eq.~(\ref{eq:g1-}) breaks down, 
so that the non-adiabatic region around $\beta_-$ is as large as possible 
without the region around $\beta_+$ being non-adiabatic as well.  
Thus, for a given axion parameters $g,~m_a$ satisfying Eq.~(\ref{eq:gm}),
the maximum conversion occurs for the photon energy between
\be
\omega \sim 0.3\, g_9^{1/2} m_5^2 (\beta_sR_6^3)^{1/2}~{\rm keV}
\ee
and
\be 
\omega \sim 2\, g_9^{5/2}\! (\beta_s R_6^3)^{1/2} ~ {\rm keV}.
\label{eq:max_conv}
\ee
This nonmonotonic behavior of $P_{\rm conv}$ is shown in 
Fig.~\ref{fig:evodat}. As noted before, for $(g/m_a)<(g/m_a)_\ast$,
no ``strong mixing'' region exists in the photon's path, 
the net conversion probability will be small, regardless of whether 
condition (\ref{eq:gamma}) is satisfied or not (see Fig.~\ref{fig:evo4}).

\subsection{Effect on Radiation Spectrum and Polarization fron Magnetic 
Neutron Stars}

\begin{figure}
\includegraphics[height=10cm]{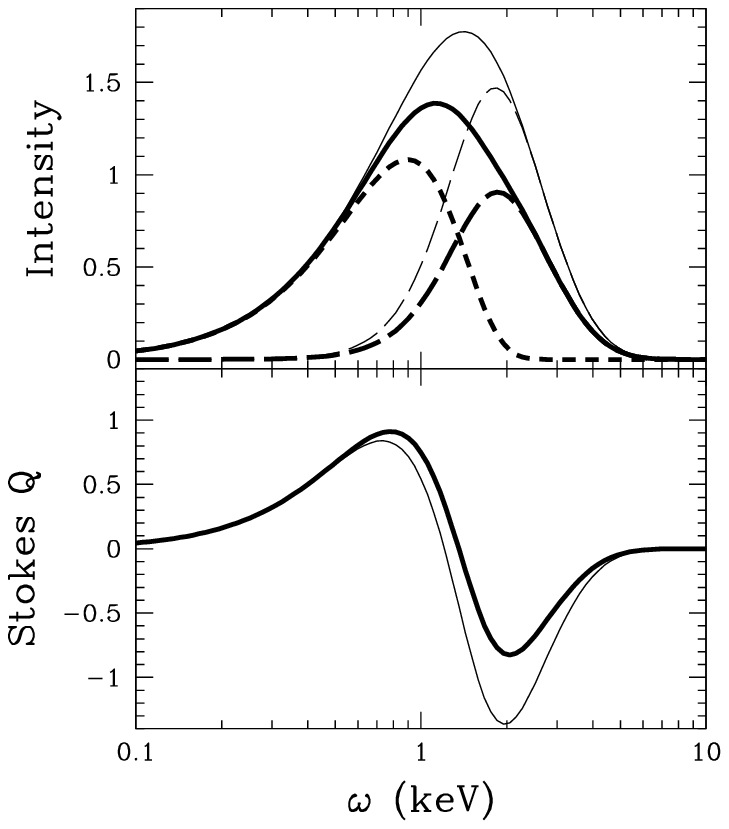}
\caption{Spectrum (upper panel) and Stokes parameter $Q$ (lower panel)
of a magnetized neutron star with surface magnetic field $B_s=10^{13}$~G
and temperature $T_s=0.5$~keV. In both panels, 
the heavy lines show the results including 
photon-axion coupling with the parameters $m_a=10^{-5}$~eV and $g=2\times 
10^{-9}~\GeV1$, while the light lines neglect photon-axion coupling. 
In the upper panel, the solid lines show the 
total radiation intensity, the short-dashed line shows the 
$\perp$-mode intensity and the long-dashed line shows 
the $\parallel$-mode intensity.
\label{fig:spec}}
\end{figure}

To illustrate the effect of photon-axion coupling on the radiation
from magnetized neutron stars (NSs), we consider an example of a NS
with surface magnetic field $B=10^{13}$~G and surface temperature
$T_s=0.5$~keV (see Fig.~\ref{fig:spec}).  As shown in
Refs.~\cite{laiho03,vanadel06}, due to the vacuum resonance effect
\cite{laiho02,laiho03a} occurring near the NS surface (but outside the
photosphere) (see also Sect.~\ref{sec:plasma} below), the surface
radiation from such a NS is dominated by the extraordinary mode (or
$\perp$ mode) for $\omega\lo 1-2$~keV and by the ordinary mode
($\parallel$ mode) for $\omega\go 1-2$~keV. In our calculation leading
to Fig.~\ref{fig:spec}, we assume (for simplicity) that the total
intensity near the NS surface is given by a blackbody with temperature
$T_s$ (see, e.g., \cite{holai03,vanadel06} for more accurate
atmosphere model spectra), and we calculate the intensities of the two
photon modes ($I_\perp$ and $I_\parallel$) outside the vacuum
resonance layer using the method described in \cite{laiho02}.  With no
photon-axion coupling, the observed photon intensity is
$I_\perp+I_\parallel$, and Stokes parameter $Q$ is
$I_\perp-I_\parallel$ (Note that the Stokes parameter depends on
the orientation of the axis used in measuring linear polarization, but
one can always choose the appropriate axis so that $U=0$;
Refs.~\cite{laiho03,vanadel06} give a more detailed description of the
phase-dependent polarization signals).  With nonzero photon-axion
coupling, as the photon propagates from the stellar surface to the
observer, the $\parallel$-mode intensity will be reduced to
$I_\parallel^\infty=(1-P_{\rm conv})I_\parallel$, where $P_{\rm conv}$
is the photon-axion conversion probability calculated before
(Fig.~\ref{fig:evodat}), while the $\perp$-mode intensity is
unchanged.  Thus the observed total radiation intensity is
$I_\perp+I_\parallel^\infty$, and the Stokes Q parameter is
$I_\perp-I_\parallel^\infty$.

For NSs with $B_s\go 7\times 10^{14}$~G, the radiation from the NS surface
is dominated by the $\perp$-mode, which is unaffected by any further 
photon-axion coupling as the photon propagates to the observer. Thus,
for such NSs, the axion would have a much smaller effect on the observed
NS radiation than depicted in Fig.~\ref{fig:spec}.

\section{Axion-Photon Propagation in an Inhomogeneous Plasma}
\label{sec:plasma}

In this section we consider the effect of photon-axion coupling on the
photon propagation in the atmospheric plasma of a magnetized NS. Here
the magnetic field is constant, but the plasma density varies on the
scale of centimeters (the atmosphere scale height at $T_s\sim
10^6$~K).

\subsection{Vacuum resonance}
\label{subsec:vacuum}

We first consider wave propagation in a magnetized plasma, including
the vacuum polarization effect, but neglecting axion-photon 
coupling~\cite{laiho02,laiho03,laiho03a,gnedin78,
meszaros79,pavlov79}.
In the weak-dispersion approximation (i.e., the index of refraction 
of the photon is very close to unity),
the electromagnetic wave equation in the plasma including vacuum
polarization (but no axion) is given by Eq.(30) of
\cite{laiho03}. Using the sign convention adopted for this paper, we
have 
\be i{d\over dz}\left(\begin{array}{c}E_\parallel\\
    E_\perp\end{array}\right) ={\omega\over 2} \left(\begin{array}{cc}
    2+\sigma_{11} & \sigma_{12} \\
    \sigma_{21} & 2+\sigma_{22}\end{array} \right)
\left(\begin{array}{c}E_\parallel \\ E_\perp \end{array}\right).
\label{eq:evop}\ee
Neglecting the protons (i.e. setting the proton mass $m_p\rightarrow \infty$) 
and damping terms in the dielectric tensors 
(more general expressions are given in \cite{laiho03}), 
the matrix elements are given by
\ba
&&\sigma_{11}=(q-v_e)\sin^2\theta-{v_e\over 1-u_e}\cos^2\theta,\\
&&\sigma_{22}=-m\sin^2\theta-{v_e\over 1-u_e},\label{eq:sigma22}\\
&&\sigma_{12}=-\sigma_{21}=i{v_eu_e^{1/2}\over 1-u_e}\cos\theta.
\label{eq:sigma12}
\ea
Here $u_e=\omega_{ce}^2/\omega^2$, with $\omega_{ce}=eB/(m_ec)=m_ec^2 b$
the electron cyclotron (angular) frequency, and
$v_e=\omega_{pe}^2/\omega^2$, with
$\omega_{pe}= (4\pi n_e e^2/m_e)^{1/2}=28.71\,
(Y_e\rho_1)^{1/2}$~eV the electron plasma (angular) frequency, 
where $n_e=Y_e\rho/m_p$ is the electron density, $\rho$ is
the mass density, $Y_e$ is the electron fraction and $\rho_1=\rho/(1~{\rm g~
cm}^{-3})$. The quantities $q,m$ are the vacuum polarization parameters:
$q$ is given by Eq.~(\ref{eq:q}) and \cite{potekhin04}
\be
m=-{4\alpha\over 45\pi}\,b^2{\hat m},\quad
{\hat m}={1\over 1+0.72 b^{5/4}+(4/15)b^2}.
\ee
The weak-dispersion approximation adopted in Eq.~(\ref{eq:evop}) is valid 
valid for $v_e\ll 1$ and $b\ll 3\pi/\alpha\sim 10^3$.

With $E_{\parallel,\perp}\propto e^{-ikz}$, Eq.~(\ref{eq:evop})
determines two normal modes. For $u_e\gg 1$, the modes are almost
linearly polarized, except at the ``vacuum
resonance''~\cite{gnedin78,meszaros79,pavlov79,laiho02,laiho03a},
which occurs at $\sigma_{11}=\sigma_{22}$, or \be v_e=(q+m){u_e-1\over
  u_e}, \ee which for $u_e\gg 1$, becomes $v_e=q+m$. For a given
photon energy, the resonance density is 
\be 
\rho_{\rm res}({\rm
  vacuum})=0.964\,Y_e^{-1}(B_{14}\omega_1)^2f_B^{-2} ~{\rm g~cm}^{-3},
\ee 
where $B_{14}=B/(10^{14}~{\rm G})$, $\omega_1=\omega/(1~{\rm keV})$,
and $f_B=[(\alpha b^2/15\pi)/(q+m)]^{1/2}$ is a slow varying function
of $B$ and is of order unity ($f_B\simeq 1$ for $b\lo 1$ and $f_B\lo
4$ for $b\lo 100$; we will use $f_B=1$ in remainder of this paper).
At the vacuum resonance, the plasma and vacuum polarization effects
are comparable, and both modes are circularly polarized.

The evolution of photon modes around the vacuum resonance in an
inhomogeneous NS atmosphere (varying density but constant magnetic
field) has been studied before~\cite{laiho02,laiho03}.  The key result
is that away from the resonance, the mode evolution is highly
adiabatic. At the resonance, the mode evolution depends on the
adiabaticity parameter
\be
\gamma_{\rm res}=(\omega/\omega_{\rm ad})^3,
\label{eq:gammares1}\ee
with
\be
\omega_{\rm ad}=2.55\,(f_B\,\tan\theta)^{2/3}H_1^{-1/3}~{\rm keV},
\label{eq:omegaad}
\ee
where 
$H=\rho/|\rho'|$
is the density scale height along the ray (evaluated at the resonance
point), and $H_1=H/(1~{\rm cm})$.  For an ionized hydrogen atmosphere,
$H=\rho/|\rho'|
=2kT/(m_pg_*\cos\theta_{kg})=1.65\,T_6/(g_{*,14}\cos\theta_{kg})$~cm,
where $T=10^6T_6$ is the temperature, $g_*=10^{14}g_{*,14}$~cm~s$^{-2}$ is
the gravitational acceleration, and $\theta_{kg}$ is the angle between
the ray and the NS surface normal.  Since nonadiabaticity can only
occur at the vacuum resonance, we can use the Landau-Zener formula to
calculate the nonadiabatic jump probability between normal modes 
\cite{kuo89}:
\be
P_{\rm jump}=e^{-\pi\gamma_{\rm res}/2}.
\label{eq:pjumpvac}\ee
Thus, for $\gamma_{\rm res}\gg 1$, the evolution is adiabatic, and the
photon can convert from the $\parallel$-mode (ordinary mode) to the
$\perp$-mode (extraordinary mode) and vice versa.  Various
implications of the vacuum resonance phenomenon for NS surface
radiation spectrum and polarization are studied in
Refs.~\cite{laiho02,laiho03,holai03,laiho03a,vanadel06}.

\subsection{Axion-Photon Resonance}
\label{subsec:photon-a}

We now consider the effect of photon-axion coupling on photon
propagation in an inhomogeneous magnetized plasma.  Combining
Eqs.~(\ref{eq:evol}) and (\ref{eq:evop}), we find that in the
weak-dispersion approximation, the evolution equation for the photon
(with two polarization components, $E_\parallel$ and $E_\perp$) and
axion takes the form
\be
i{d\over dz}\,\Phi
=\left(\begin{array}{ccc}
\omega+\Delta_a & \Delta_M & 0\\
\Delta_M & \omega+\Delta_\parallel+\Delta_p & \sigma_{12}\omega/2\\
0 & \sigma_{21}\omega/2 &\omega+\sigma_{22}\omega/2\end{array} \right)\Phi,
\label{eq:evolu}\ee
where
\be
\Phi=
\left(\begin{array}{c}a \\ E_\parallel \\ E_\perp \end{array}\right).
\ee
In Eq.~(\ref{eq:evolu}), 
$\Delta_a$, $\Delta_\parallel$ and $\Delta_M$ are the same
as in Eq.~(\ref{eq:evol}), $\sigma_{22}$,~$\sigma_{12}$,~$\sigma_{21}$ are 
given by Eqs.~(\ref{eq:sigma22})-(\ref{eq:sigma12}), and 
\be
\Delta_p=-{\omega v_e\over 2}F,\qquad {\rm with}~~~
F={\cos^2\theta\over 1-u_e}+\sin^2\theta.
\ee
Obviously, in vaccum ($v_e=0$),
Eq.~(\ref{eq:evolu}) reduces to Eq.~(\ref{eq:evol}); 
with no photon-axion coupling ($\Delta_M=0$), it reduces to 
Eq.~(\ref{eq:evop}).

The {\it axion-photon resonance} 
occurs when the first and second diagonal matrix
elements in Eq.~(\ref{eq:evolu}) are equal, i.e., 
$\Delta_a=\Delta_\parallel+\Delta_p$, or
\be
v_e={q\sin^2\theta+(m_a^2/\omega^2)\over F}\qquad 
({\rm axion-photon~resonance}).
\label{eq:veaxion}\ee
Equations (\ref{eq:evolu})-(\ref{eq:veaxion}) are general, and are valid
for different magnetic field regimes, such as those found in NSs, white
dwarfs and the Sun.

We now consider magnetic field strengths relevant to NSs and assume
$u_e\gg 1$. We consider a typical angle $\theta$
that is not too close to $0^\circ$ or $180^\circ$. For the
parameters (axion mass, photon energy and field strength) of interest,
$q\sin^2\theta\gg m_a^2/\omega^2$ (or $\Delta_\parallel\gg 
|\Delta_a|$) is well satisfied. Then the axion-photon resonance
is at 
\be
v_e=q\qquad ({\rm axion-photon~resonance}).
\label{eq:veq}\ee
Note that the vacuum resonance is at $v_e=q+m$.
The two resonances are well separated, and we can treat them
independently --- we justify this in Sect.~\ref{subsec:3state}.

The vacuum resonance has been discussed in Sect.~\ref{subsec:vacuum}.
We now consider the axion-photon resonance. Neglecting the $E_\perp$ 
component of the photon, the basic equation takes the form
\be
i{d\over dz}\left(\begin{array}{c}a\\E_\parallel\end{array}\right)
=\left(\begin{array}{cc}
\omega+\Delta_a & \Delta_M \\
\Delta_M & \omega+\Delta_\parallel+\Delta_p \end{array} \right)
\left(\begin{array}{c}a \\ E_\parallel \end{array}\right).
\label{eq:evolu1}\ee
Similar to Eqs.~(\ref{eq:delk}) and (\ref{eq:tan}), we define
$\Delta k$ and the mixing angle $\theta_m$ via
\ba
&&\Delta k=2\left[(\Delta_a-\Delta_\parallel-\Delta_p)^2/4+\Delta_M^2
\right]^{1/2},\label{eq:delk2}\\
&&\tan 2\theta_m={\Delta_M\over (\Delta_a-\Delta_\parallel-\Delta_p)/2}.
\label{eq:tan2}\ea
The eigenvalues and eigenvectors of Eq.~(\ref{eq:evolu1}) are
\ba
&&k_\pm=\omega+{\Delta_\parallel+\Delta_a+\Delta_p\over 2}\pm
{\Delta k\over 2},\\
&&\left(\!\begin{array}{c} a \\ E_\parallel\end{array}\!\right)_+
\!\!\!=\!\left(\!
\begin{array}{c}\cos\theta_m\\ \sin\theta_m\end{array}\!\!\right),~
\left(\!\begin{array}{c}a \\ E_\parallel\end{array}\!\right)_-
\!\!\!=\!\left(\!\begin{array}{c}
-\sin\theta_m\\ \cos\theta_m\end{array}\!\!\right).
\ea
If we plot $k_\pm$ as a function of the plasma density, we find
the two eigenstates tend to ``cross'' at $\Delta_a=\Delta_\parallel+\Delta_p$,
where $\theta_m=\pm 45^\circ$ --- this is the axion-photon resonance.
Similar to Sect.~\ref{subsec:evol}, we find that the mode evolution
depends on the adiabaticity ratio $\gamma=\Delta k/(2|\theta_m'|)$.
Assuming that the variation comes only from density 
(while the magnetic field is constant), we have
\be
\theta_m'={\Delta_p\over 4\Delta_M}{\rho'\over\rho}\sin^2 2\theta_m,
\ee
Thus the adiabaticity ratio is
\ba
&& \gamma=\left|{\Delta k/2\over\theta_m'}\right|=
{4\Delta_M^2 H\over |\Delta_p\sin^32\theta_m|}\nonumber\\
&&\quad ={4\Delta_M^2 H\over \Delta_\parallel} 
{\rho_{\rm res}/\rho\over |\sin^3 2\theta_m|},
\ea
where in the last equality we have used $\Delta_\parallel\gg |\Delta_a|$, and
$\rho_{\rm res}$ is the axion-photon resonance density (for a given
photon energy $\omega$):
\be
\rho_{\rm res}=2.25\,Y_e^{-1}(B_{14}\omega_1)^2{\hat q}~{\rm g~cm}^{-3}.
\ee
At the axion-photon resonance $\rho=\rho_{\rm res}$, 
$|\theta_m|=\pi/4$, we find 
\be 
\gamma_{\rm res}={4\Delta_M^2H\over\Delta_\parallel}
=1.915 \times 10^{-8} \,{g_{9}^2H_1\over \omega_1\hat q},
\label{eq:gammmares2}\ee
Note that
\be
\sin 2\theta_m={M\over \left[(1-v_e/q)^2\sin^4\theta/16+M^2
\right]^{1/2}},
\ee
with 
\be
M={\Delta_M\over \omega q}=3.614\times 10^{-7}
\,{g_{9}\sin\theta\over b\omega_1\hat q}.
\ee
For $|1-v_e/q|\gg 4M/\sin^2\theta$, we have
\be
\gamma\simeq 6.34\times 10^9{H_1 \beta^3\omega_1^2 {\hat q}^2
\over g_{9}}{(1-v_e/q)^3\over v_e/q}.
\ee
Thus, away from the resonance, the adiabatic condition ($\gamma\gg 1$)
is well satisfied, and nonadiabaticity can only occur very close to
the resonance.  We can use the Landau-Zener formula to calculate the
nonadiabatic jump probability \cite{kuo89}
\be
P_{\rm jump}=e^{-\pi\gamma_{\rm res}/2}.
\label{eq:pjump}
\ee
Note that because of the level crossing, adiabatic mode evolution
($\gamma_{\rm res}\gg 1$) corresponds to a conversion of the
$\parallel$-mode photon to the axion and vice versa. Thus the
axion-photon conversion probability across the resonance is $P_{\rm
  conv}=1-P_{\rm jump}$.

Comparing Eqs.~(\ref{eq:gammares1})-(\ref{eq:omegaad}) with 
Eq.~(\ref{eq:gammmares2}), it is interesting to note that high-energy 
photons tend to be more adiabatic at the vacuum resonance and but 
less adiabatic at the axion-photon resonance than low-energy photons.

In Sect.~\ref{subsec:numer} we shall present numerical calculations
that confirm the analytical consideration given above.


\subsection{More Rigorous Treatment of Three-State Mixing}
\label{subsec:3state}

\begin{figure}
\includegraphics[height=7cm]{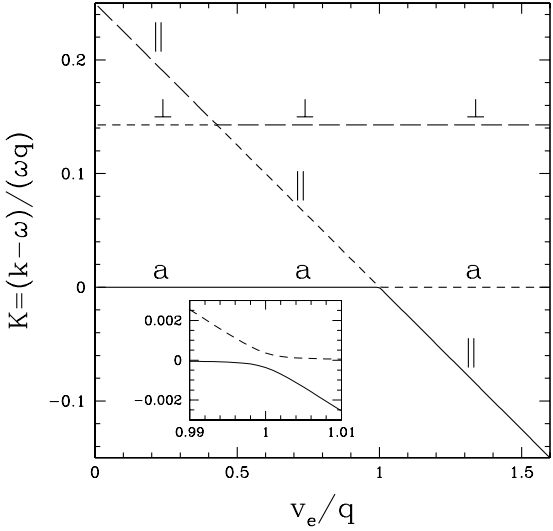}
\caption{Eigenvalues of the axion-photon system in the 
plasma+vacuum medium as a function of the plasma density parameter
$V=v_e/q$, where $v_e=(\omega_{pe}/\omega)^2\propto \rho$.
The insert shows the blowup of the region around the axion-photon
resonance. The parameters are chosen so that $A,~M,~u_e^{-1/2}\ll 1$.
\label{fig:eigen}}
\end{figure}

Here we consider the eigenmodes of the three-state system, 
Eq.~(\ref{eq:evolu}). We will show that the vacuum resonance and the
axion-photon resonance can be considered independently of each other.

The $3\times 3$ matrix in 
Eq.~(\ref{eq:evolu}) can be written as 
(for $u_e\gg 1$)
\be
\omega\,{\bf I}+\omega q \,{\bf G},
\ee
where ${\bf I}$ is the unit $3\times 3$ unit matrix, and 
\be
{\bf G}= \left[\begin{array}{ccc}
A & M & 0\\
M & {1\over 2}(1-V)\sin^2\theta & -iV\cos\theta/(2u_e^{1/2})\\
0 & iV\cos\theta/(2u_e^{1/2}) & (2\hat m/7\hat q)\sin^2\theta
\end{array} \right],
\label{eq:Gmatrix}\ee
with
\ba
&& V\equiv v_e/q,\\
&& A\equiv \Delta_a/(\omega q)\ll 1,\\
&& M\equiv \Delta_M/(\omega q)=3.614\times 10^{-7}{g_{9}\sin\theta
\over b\omega_1 \hat q}\ll 1.
\ea
Let $(a,E_\parallel,E_\perp)\propto e^{-ikz}$, and we define dimensionless
$K$ via
\be
k=\omega+\omega q K.
\ee
Obviously, $K$ depends only on the dimensionless
parameters, $V,~A,~M,~u_e$. Numerical values for $K$ 
as a function of $V$ are shown in Fig.~\ref{fig:eigen}.

Since $M\ll 1$, $u_e^{-1/2}\ll 1$, it is clear that away from the two
resonances, the eigenvalues are simply 
\ba
&& K=A \quad ({\rm axion}),\\
&& K=(1/2)(1-V)\sin^2\theta\quad (\parallel{\rm -mode}),\\
&& K=(2\hat m/7\hat q)\sin^2\theta\quad (\perp{\rm -mode}).  \ea

We now consider the mode properties in the vicinity 
of the axion-photon resonance, $|V-1|\ll 1$. The third eigenvalue is simply 
$K=(2\hat m/7\hat q)\sin^2\theta$, describing the $\perp$-mode of photon. 
The other two modes that ``intersect'' have eigenvalues that satisfy 
$|K_\pm|\ll 1$, and are given by 
\ba
&& K_\pm={A-C\over 2}+{1-V\over 4}\sin^2\!\theta\nonumber\\
&& \qquad \pm {1\over 4}
\left[(V_{ap}-V)^2\sin^4\!\theta+(4M)^2\right]^{1/2},
\ea
where 
\be
C
\simeq {7\hat q\over 8\hat m u_e\tan^2\theta}\ll 1,
\ee
and $V_{ap}$ is the value of $V$ at which the axion-photon resonance 
occurs:
\be
V_{ap}=1-{2(A+C)\over \sin^2\theta}.
\label{eq:vap}
\ee
Note that if we set $C=0$, Eq.~(\ref{eq:vap}) reduces to 
Eq.~(\ref{eq:veaxion}). 
We see that effect of including the coupling with third mode (the $\perp$-mode)
is to shift the axion-photon resonance location by a small amount.
The eigenvectors near the resonance are
\be
\Phi_+\propto
\left(\!\begin{array}{c}\cos\theta_m\\ \sin\theta_m
\\ E_{\perp+}\end{array}\!\right),
\quad
\Phi_-\propto
\left(\!\begin{array}{c}-\sin\theta_m\\ \cos\theta_m
\\ E_{\perp-}\end{array}\!\right),
\ee
where 
\be
E_{\perp\pm}=-i{7\hat q V\cos\!\theta\over 4\hat m u_e^{1/2}\sin^2\!\theta}\,
E_{\parallel\pm},
\ee
with $|E_{\perp\pm}|\ll |E_{\parallel\pm}|$,
and
\be
\tan 2\theta_m
={4M\over (V-V_{ap})\sin^2\theta}.
\ee
Again, the above expression reduces to Eq.~(\ref{eq:tan2}) for $C=0$.
We see that except for the slight shift of the resonance location,
$\Delta K=K_+-K_-$ and $\theta_m'$ are the same as those given in
Sect.~\ref{subsec:photon-a}, where the coupling with the $\perp$-mode
is neglected.  We conclude that the analysis of the axion-photon
resonance given in Sect.~\ref{subsec:photon-a} is accurate.

We can similarly show that the vacuum resonance discussed in
Sect.~\ref{subsec:vacuum} is hardly affected by the photon-axion
coupling.

\subsection{Numerical Examples of Three-Mode Evolution}
\label{subsec:numer}

\begin{figure}
\includegraphics[height=9.5cm]{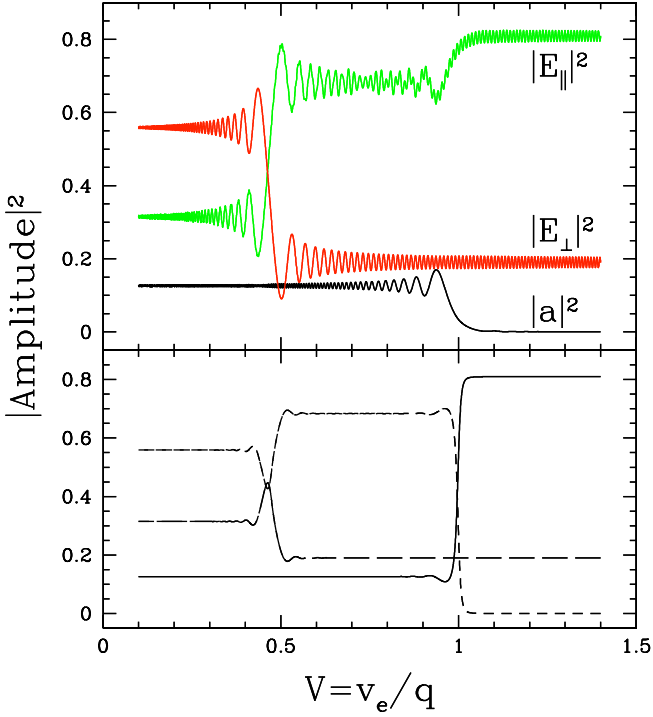}
\caption{The evolution of the three-state system in an inhomogeneous
neutron star atmosphere. The upper panel shows the square amplitudes
$|a|^2$, $|E_\parallel|^2$ and $|E_\parallel|^2$; the lower panel
shows the square amplitude of non-crossing eigenmodes, $|A_1|^2,~|A_2|^2,~
|A_3|^3$. The evolution starts in the high-density (large 
$v_e=\omega_{pe}^2/\omega^2$) region, where amplitudes are set to
$E_\parallel=0.9,~E_\perp=\sqrt{0.19}$ and $a=0$. The parameters are
$m_a=10^{-3}$~eV, $g=10^{-6}~\GeV1$, $B=10^{13}$~G, $H=5$~cm, 
$\theta=45^\circ$, and $\omega=1$~keV.
\label{fig:f10}}
\end{figure}

\begin{figure}
\includegraphics[height=9.5cm]{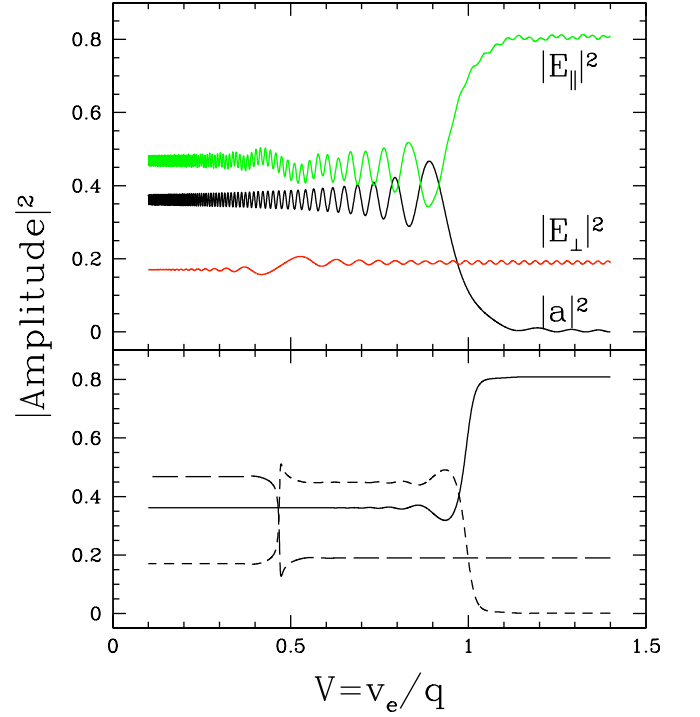}
\caption{Same as Fig.~10. except $\omega=0.3$~keV.
\label{fig:f11}}
\end{figure}

Here we consider the numerical integration of the axion-photon system
in a neutron star atmosphere, with a constant magnetic field ${\bf B}$
but varying densities. In our examples, we assume
$\rho=\rho_a\,\exp(-z/H)$, where $\rho_a$ is the density of the
axion-photon resonance, and $H$ is the density scale height along the
direction of propagation. Equation (\ref{eq:evolu}) is equivalent to
\be {d\over dV}\Phi=i\,{\omega q H\over V}\,{\bf G}\,\Phi,
\label{eq:dphi}\ee
with ${\bf G}$ given by (\ref{eq:Gmatrix}). We integrate (\ref{eq:dphi})
outward from the high-density region to the low-density region,
traversing the axion-photon resonance at $V=1$ and the vacuum resonance
at $V=1+m/q$. Fig.~\ref{fig:f10} and Fig.~\ref{fig:f11} give two
examples of the evolution, with the photon energy $\omega =1$~keV 
(Fig.~\ref{fig:f10}) and $\omega =0.3$~keV (Fig.~\ref{fig:f11}), 
respectively, both with the parameters
$m_a=10^{-3}$~eV, $g=10^{-6}~\GeV1$, $B=10^{13}$~G, $H=5$~cm, 
and $\theta=45^\circ$, and the same initial conditions at $V=1.4$:
$E_\parallel=0.9,~E_\perp=\sqrt{0.19}$ and $a=0$. 
The lower panels of Figs.~\ref{fig:f10}-\ref{fig:f11} depict
$|A_i|^2$ ($i=1,2,3$), where the mode amplitude $A_i$ is defined by
\be
\Phi=\sum_{i=1}^3 A_i\,\Phi_i,
\ee
and $\Phi_i$ are the normalized mode eigenvectors ($\Phi_i^+\Phi_i=1$)
discussed in Sect.~\ref{subsec:3state}, and $A_i=\Phi_i^+\Phi$.

The numerical results fully agree with the analytical consideration of
Sect.~\ref{subsec:3state}, and show that the two resonances can be
treated independently of each other. In particular, the mode evolution
across the resonance is described by the non-adiabatic jump
probability to within about 30\%, Eq.~(\ref{eq:pjumpvac}) (for vacuum
resonance) or (\ref{eq:pjump}) (for axion-photon resonance) for values
of $\gamma_{\rm res}$ that vary by nearly seven orders of magnitude from
$10^{-6}$ to 10.

\section{Axion-Photon Propagation in the Sun and White dwarfs}

The central core of the Sun may a source of keV axions. These axions
are the target of the CAST experiment~\cite{zioutas05}.
Could the axions oscillate/transform into photons as they
travel across the solar envelope and atmosphere?

The general axion-photon propagation equation (\ref{eq:evolu})
can be applied here, and we have $u_e\ll 1$ and $\Delta_\parallel$ 
is negligible. As in Sect.~\ref{subsec:photon-a}, 
it is adequate to consider Eq.~(\ref{eq:evolu1}), with $\Delta_p=
-\omega v_e/2=-\omega_{pe}^2/(2\omega)$. The axion-photon mixing angle
is given by [see Eq.~(\ref{eq:tan2})]
\be
\tan 2\theta_m={\Delta_M\over (\Delta_a-\Delta_p)/2}.
\ee
Since $\Delta_M/|\Delta_a|=5.92\times 10^{-8}m_{5}^{-2}g_{9}\omega_1
B_1\sin\theta$ (where $B_1$ is the field strength in units of Gauss),
we see that the mixing angle is quite small except near the resonance,
where $\Delta_p=\Delta_a$. 
The resonance occurs when $m_a=\omega_{\rm pe}$, at the density
$\rho=1.21\times 10^{-5}Y_e^{-1}m_{5}^2$~g~cm$^{-3}$. 
At the resonance, the adiabaticity parameter is 
\be
\gamma_{\rm res}
={4\Delta_M^2H\over |\Delta_a|}=4.2\times10^{-5} \,g_{9}B_1\sin\theta\,
{\Delta_M\over|\Delta_a|}{H\over R_\odot},
\ee
where $R_\odot$ is the solar radius and $H$ is the density scale
height. Clearly, $\gamma_{\rm res}\ll 1$ and no resonant conversion of
axion to photon is expected around the resonance.

In the case of magnetic white dwarfs, with $B$ in the range of $10^3$-$10^9$~G,
we still consider $u_e\ll 1$. The axion-photon resonance is at 
$v_e=q\sin^2\theta+(m_a/\omega)^2$, or the density 
\be
\rho_{\rm res}=1.21\times 10^{-5}Y_e^{-1}m_{5}^2\,
\left(1+\Delta_\parallel/|\Delta_a|\right)\,{\rm g}~{\rm cm}^{-3},
\label{eq:rhores}\ee
where $\Delta_\parallel/|\Delta_a|=1.855 \times 10^3 \,\omega_1^2B_9^2m_{-5}^{-2}
\sin^2\theta$, and $B_9=B/(10^9~{\rm G})$. The adiabaticity parameter at 
the resonance is
\ba
&&\gamma_{\rm res}
={4\Delta_M^2H\over |\Delta_a|(1+|\Delta_\parallel/|\Delta_a|)}\nonumber\\
&&=2.49\times 10^5{\omega_1 \left(m_{5}^{-1}g_{9}B_9\sin\theta
\right)^2\over 1+|\Delta_\parallel/|\Delta_a|}
\left({H\over 10^{-2}R_\odot}\right),
\ea
where we have scaled the density scale height $H$ to the typical radius of the
star, $10^{-2}R_\odot$. Thus adiabatic resonant axion-photon conversion 
is a possibility. 
For a given set of axion parameters, we expect that the flux of 
photons satisfying $\gamma_{\rm res}\gg 1$ is reduced by a factor of 2, 
and significant linear polarization will be present. Observations of 
the spectra and polarizations of magnetic white dwarfs would provide
constraint on the axion parameters. To carry out such constraint
quantitatively, it is necessary to model the low-density 
region [see Eq.~(\ref{eq:rhores})] of the white dwarf atmosphere,
a task beyond the scope of our paper. 

The evolution of the axion-photon system 
in the vacuum region outside a white dwarf 
with a varying magnetic field is similar to the neutron star case studied in 
Sect.~\ref{sec:vacuum}.

\section{Discussion}

In this paper we have studied the propagation of the axion-photon system 
in the atmospheres and the near vicinity of magnetic stars.
Because of their strong magnetic fields, we have focused on neutron stars,
but our main results/methods can be similarly applied to other
astrophysical bodies. Our study goes beyond previous 
work~\cite{raffelt88} in that we quantify the various parameter regimes
for which axion-photon conversion is important, we calculate
the conversion probabilities for propagation in varying magnetic fields
and varying plasma densities, and we present examples to illustrate 
how axion-photon coupling may affect observed photon spectra
and polarizations.

The axion-photon resonance (where maximum mixing occurs) is always
present in magnetized neutron star atmospheres/magnetospheres.  This
resonance [see Eq.~(\ref{eq:veq})] is at a higher density than the
previously-studied vacuum resonance. Complete axion-photon conversion
is possible only when the adiabatic condition [$\gamma_{\rm res}\gg
1$; see Eq.~(\ref{eq:gammmares2})] is satisfied. A high axion-photon
coupling strength, gradual density gradient and lower photon energy
tend to make such conversion a possibility.  Note that in this paper
we have treated the plasma dielectric tensor in the cold-plasma
approximation. For low-energy photons (e.g. optical), the resonance occurs
in the neutron star's magnetosphere and the cold plasma treatment is 
no longer valid. It would be straightforward to to generalize our results
to more general plasma dielectric tensors.

Even without the axion-photon resonance, partial axion-photon conversion may
take place during the propagation in the vacuum region with spatially 
varying magnetic fields (see Sect.~II).

Applying our result to the axions produced at the center of the Sun, 
we find that there is no possibility for appreciable axion-photon 
conversion during propagation (see Sect.~IV). 
Thus the nondetection of axions in the CAST experiment cannot be
explained by such oscillation effect.
Our analysis also shows that with the axion parameters allowed by the 
PVLAS experiment, significant photon-axion resonant
conversion is possible in highly magnetized white dwarfs. This may 
produce interesting spectral and polarization signatures in the observed
radiation from the white dwarf.

\bigskip
\begin{acknowledgments}
  This work was supported in part by NSF grant AST 0307252 and
  SAO Grant TM6-7004X (DL), as well as a
  Discovery Grant from NSERC (JSH).  
%
 This work made use of NASA's Astrophysics Data
  System.  The authors were visitors at the Pacific Institute of
  Theoretical Physics during the nascent stages of this research.
\end{acknowledgments}



\end{document}